\documentclass[twocolumn,usenatbib]{mn2e}
\usepackage{amsmath}
\usepackage{epsfig}
\usepackage{capt-of}

\begin{document}

\title{An Infinite Family of Generalized Kalnajs Disks}

\author[Guillermo A. Gonz\'alez and Jerson I. Reina]{Guillermo A.
Gonz\'alez\thanks{E-mail: guillego@uis.edu.co} and Jerson I.
Reina\thanks{E-mail: jeivreme@hotmail.com}\\ Escuela de F\'isica, Universidad
Industrial de Santander, A. A. 678, Bucaramanga, Colombia}

\maketitle

\begin{abstract}
An infinite family of axially symmetric thin disks of finite radius is
presented. The family of disks is obtained by means of a method developed by
Hunter and contains, as its first member, the Kalnajs disk. The surface
densities of the disks present a maximum at the center of the disk and then 
decrease smoothly to zero at the edge, in such a way that the mass distribution
of the higher members of the family is more concentrated at the center. The
first member of the family have a circular velocity proportional to the radius,
representing thus a uniformly rotating disk. On the other hand, the circular
velocities of the other members of the family increases from a value of zero at
the center of the disks until a maximum and then decreases smoothly until a
finite value at the edge of the disks, in such a way that for the higher
members of the family the maximum value of the circular velocity is attained
nearest the center of the disks.
\end{abstract}


\section{Introduction}

A fact usually assumed in astrophysics is that the main part of the mass of a
typical spiral galaxy is concentrated in a thin disk (\cite{BT}). Accordingly,
the obtention of the gravitational potential generated by an idealized thin
disk is a problem of great astrophysical relevance and so, through the years,
different approaches has been used to obtain such kind of thin disk models.

Wyse and Mayall (\citeyear{WM}) studied thin disks by superposing an infinite
family of elementary disks of different radii. Brandt (\citeyear{BR}) and
Brandt and Belton (\citeyear{BB})  constructed flat galaxy disks by the
flattening of a distribution of matter whose surface of equal density were
similar spheroids. A simple potential-density pair for a thin disk model was
introduced by Kuzmin (\citeyear{KUZ}) and then rederived by Toomre
(\citeyear{T1,T2}) as the first member of a generalized family of models.

The Toomre models are obtained by solving the Laplace equation in cylindrical
coordinates subject to appropriated boundary conditions on the disk and at
infinity. The Kuzmin and Toomre models of thin disks, although they have
surface densities and rotation curves with remarkable properties, represent
disks of infinite extension and thus they are rather poor flat galaxy models.
Accordingly, in order to obtain more realistic  models of flat galaxies, is
better to consider methods that permit the obtention of finite thin disk
models. 

A simple method to obtain the surface density, the gravitational potential and
the rotation curve of thin disks of finite radius was developed by \cite{HUN1}.
The Hunter method is based in the obtention of solutions of Laplace equation in
terms of oblate spheroidal coordinates, which are ideally suited to the study
of flat disks of finite extension. By superposition of solutions of Laplace
equation, expressions for the surface density of the disks, the gravitational
potential and its rotational velocity can be obtained as series of elementary
functions. 

The simplest example of a thin disk obtained by means of the Hunter method is
the well known Kalnajs disk (\cite{KAL}), which can also be obtained by
flattening a uniformly rotating spheroid (\cite{WM,BR,BB}). The Kalnajs disk
have a well behaved surface density and represents a uniformly rotating disk,
so that its circular velocity is proportional to the radius, and its stability
properties  have been extensively studied (see, for instance, Hunter
(\citeyear{HUN1,HUN2}), \cite{KAL} and \cite{K-A}).

In this paper we use the Hunter method in order to obtain an infinite family of
thin disks of finite radius. We particularize the Hunter general model by
considering a family of thin disks with a well behaved surface mass density. We
will require that the surface density be a monotonically decreasing function of
the radius, with a maximum at the center of the disk and vanishing at the edge,
in such a way that the mass distribution of the  higher members of the family
be more concentrated  at the center.

The paper is organized as follows. In Sec. 2 we present a summary of the Hunter
method used to obtain the thin disk models of finite radius and also we obtain
the general expressions for the gravitational potential, the surface density
and the circular velocity. In the next section, Sec. 3, we present the
particular family of models obtained by imposing the required behavior of the
surface densities and then, in Sec. 4, we analyze its physical behavior.
Finally, in Sec. 5, we summarize our main results.

\section{General Finite Thin Disk Models}

In order to obtain finite axially symmetric thin disk models, we need to find
solutions of the Laplace equation that represents the outer potential of a thin
disklike source. According with this, we need to solve the Laplace equation for
an axially symmetric potential,
\begin{equation}
\Phi_{,RR} + \frac{\Phi_{,R}}{R} + \Phi_{,zz} = 0,\label{eq:laplace}
\end{equation}
where $(R,\phi,z)$ are the usual cylindrical coordinates. We will suppose that,
besides the axial symmetry, the gravitational potential has symmetry of
reflection with respect to the plane $z = 0$,
\begin{equation}
\Phi(R,z) = \Phi({R},-z), \label{eq:con1}
\end{equation}
so that the normal derivative of the potential, $\partial\Phi/\partial z$,
satisfies the relation
\begin{equation}
\frac{\partial\Phi}{\partial z}(R,-z) = -\frac{\partial\Phi}{\partial z}(R,z),
\label{eq:con2}
\end{equation}
in agreement with the attractive character of the gravitational field. We also
assume that $\partial \Phi/\partial z$ do not vanishes on the plane $z=0$, in
order to have a thin distribution of matter that represents the disk. 

Given a potential $\Phi (R,z)$ with the above properties, the density $\Sigma
(R)$ of the surface distribution of matter can be obtained using the Gauss law
(\citet{BT}). So, using the equation (\ref{eq:con2}), we obtain
\begin{equation}
\Sigma(R) = \frac{1}{2\pi G}\left[\frac{\partial\Phi}{\partial z}\right]_{z =
0^{+}}.\label{eq:sigma}
\end{equation}
Now, in order to have a surface density corresponding to a finite disklike
distribution of matter, we impose the boundary conditions
\begin{subequations}\begin{align}
\frac{\partial \Phi}{\partial z}(R,0^{+}) \neq 0 ; \qquad R\leq a, \label{eq:con3} \\
\frac{\partial \Phi}{\partial z}(R,0^{+}) = 0 ; \qquad R > a,\label{eq:con4}
\end{align}\end{subequations}
so that the matter distribution is restricted to the disk $z=0$, $0\leq R\leq
a$.

We introduce now the oblate spheroidal coordinates, whose symmetry adapts in a
natural way to the geometry of the model. This coordinates are related to the
usual cylindrical coordinates by the relation (\citet{MF}),
\begin{subequations}\begin{align}
R &= a \sqrt{(1+\xi^{2}) (1-\eta^{2})}, \\
z &= a \xi \eta, 
\end{align}\end{subequations}
where $0 \leq \xi<\infty$ and $-1\leq \eta< 1$. The disk has the coordinates
$\xi = 0$, $0 \leq \eta^2 < 1$. On crossing the disk, $\eta$ changes sign but
does not change in absolute value. This singular behavior of the coordinate
$\eta$ implies that an even function of $\eta$ is a continuous function
everywhere but has a discontinuous $\eta$ derivative at the disk.

In terms of the oblate spheroidal coordinates, the Laplace equation can be
written as
\begin{equation}
[(1 + \xi^2) \Phi_{,\xi}]_{,\xi} + [(1 - \eta^2) \Phi_{,\eta}]_{,\eta},
\end{equation}
and we need to find solutions that be even functions of $\eta$ and with the
boundary conditions
\begin{subequations}\begin{align}
\Phi_{,\xi} (0,\eta) &= F (\eta ) , \label{eq:bc1} \\
\Phi_{,\eta} (\xi,0) &= 0 , \label{eq:bc2}
\end{align}\end{subequations}
where $F(\eta)$ is an even function which can be expanded in a series of
Legendre polynomials in the interval $- 1 \leq \eta \leq 1$ (\cite{BAT}).

According with this, the Newtonian gravitational potential for the exterior of
a finite thin disk with an axially symmetric matter density can be written as
(\citet{BAT}),
\begin{equation}
\Phi(\xi,\eta) = - \sum_{n=0}^{\infty} C_{2n} q_{2n}(\xi) P_{2n}(\eta),
\label{eq:potencial}
\end{equation}
where $C_{2n}$ are arbitrary constants, $P_{2n}(\eta)$ and $q_{2n}(\xi)=
i^{2n+1}Q_{2n}(i\xi)$ are the usual Legendre polynomials and the Legendre
functions of second kind, respectively. With this general
solution for the gravitational potential, the surface matter density is given
by
\begin{equation}
\Sigma(R) = \frac{1}{2\pi a G \eta}\sum_{n = 0}^{\infty} C_{2n} (2n+1)
q_{2n+1}(0) P_{2n}(\eta) \label{eq:4.10}
\end{equation}
and, as we will shown later, the arbitrary constants $C_{2n}$ must be chosen
properly so that the surface density presents a physically reasonable behavior.

Besides the matter density, another quantity commonly used to characterize
galactic matter distributions is the circular velocity $V(R)$, also called the
rotation curve, defined as the tangential velocity of the stars in circular
orbits around the center. Now, given  $\Phi(R,z)$, we can easily evaluate $V$
through the relation
\begin{equation}
V^{2} =  R \left[ \frac{\partial\Phi}{\partial R}\right]_{z=0},
\label{eq:1.22}
\end{equation}
in such a way that, by using (\ref{eq:potencial}), we obtain 
\begin{equation}
V^{2}(R) = \frac{R^2}{(a^2 - R^2)^{1/2}} \sum_{n=0}^{\infty} C_{2n}
q_{2n}(0) P'_{2n}(\eta). \label{eq:velocidad}
\end{equation}
for the circular velocity.

\section{The Generalized Kalnajs Disks}

We will now particularize the above general model by considering a family of
finite thin disk with a well behaved surface mass density. We will require that
the surface density will be a monotonically decreasing function of the radius,
with a maximum at the center of the disk and vanishing at the edge. In order to
do this, we impose the conditions
\begin{align}
\Sigma(a) &= 0, \label{eq:4.13} \\
\Sigma(0) &= \Sigma_{max}, \label{eq:4.14}
\end{align}
and we also require that
\begin{equation}
M = {2\pi} \int_{0}^{a} \Sigma(R) R dR ,\label{eq:4.12}
\end{equation}
where $M$ is the total mass of the disk.

Now, by using the boundary condition (\ref{eq:bc1}), the surface density can be
written in the form
\begin{equation}
\Sigma(R) = \frac{F(\eta)}{2\pi a G \eta}, \label{eq:4.11}
\end{equation}
where $F (\eta)$ is an even function of $\eta$, monotonically increasing at the
interval $0 \leq \eta \leq 1$, and such that
\begin{equation}
\lim_{\eta \to 0} \frac{F(\eta)}{\eta} = 0 . \label{eq:limit}
\end{equation}
Furthermore, we must to impose the condition
\begin{equation}
\int_{0}^{1} F (\eta) d\eta = \frac{M G}{a} ,
\end{equation}
in agreement with (\ref{eq:4.12}).
 
A simple function $F(\eta)$ that agrees with all the above requirements was
given by \citet{LO} and can be written as
\begin{equation}
F(\eta) = (2m+1)\frac{MG}{a} \eta^{2m}, \label{eq:4.15a}
\end{equation}
where, in order to fulfill the condition (\ref{eq:limit}), we must take $m \geq
1$. With this particular choice of $F(\eta)$ we obtain an infinite family of
finite disks with surface mass densities given by
\begin{equation}
\Sigma_{m}(R) = \frac{(2m+1)M}{2\pi a^{2}} \left[ 1 - \frac{R^{2}}{a^{2}}
\right]^{m-1/2}. \label{eq:4.20}
\end{equation}
As we can easily see, the disk with $m = 1$ corresponds to the well known
Kalnajs disk (\cite{KAL}). Accordingly, this family of finite thin disks can
then be considered as a generalization of the Kalnajs disk

Now, from the equation (\ref{eq:4.10}), the function $F(\eta)$ can be
written as
\begin{equation}
F(\eta)= \sum_{n=0}^{\infty}K_{2n}P_{2n}(\eta), \label{eq:4.15}
\end{equation}
with
\begin{equation}
K_{2n}=  (2n+1) q_{2n+1}(0) C_{2n}. \label{eq:4.16}
\end{equation}
The coefficients $K_{2n}$ are founded, by using the orthogonality property of
the Legendre polynomials, through the expression
\begin{equation}
K_{2n} = \frac{4n+1}{2} \int_{-1}^{1} F(\eta) P_{2n}(\eta) d\eta
.\label{eq:4.17}
\end{equation}
The above equation can be expressed as (\citet{BAT2})
$$
K_{2n} = \frac{M G}{2a} \left[ \frac{ \pi^{1/2}  (4n+1) (2m+1)
\Gamma(2m+1)}{2^{2m} \Gamma(1 + m - n) \Gamma(m + n + \frac{3}{2})} \right], 
$$
so that, using the gamma function properties, we obtain:
$$
C_{2n}= \frac{M G}{2a} \left[ \frac{\pi^{1/2} (4n+1) (2m+1)!}{2^{2m} (2n+1) (m
- n)! \Gamma(m + n + \frac{3}{2} ) q_{2n+1}(0)} \right],
$$
for $n \leq m$ and $C_{2n} = 0$ for $n > m$.

\section{Behavior of the Models}

With the above values of the $C_{2n}$ we can compute the different physical
quantities that characterize the behavior of the models. So, for instance, the
gravitational potential of the first three members of the family are given by
\begin{subequations}\begin{align}
\Phi_{1}(\xi,\eta) &= - \frac{MG}{a} [ \cot^{-1}\xi  + A (3\eta^{2} - 1)],
\label{eq:4.22}   \\
\Phi_{2}(\xi,\eta) &= - \frac{MG}{a} [ \cot^{-1} \xi + \frac{10 A}{7}
(3\eta^{2} - 1) \nonumber \\
 &  \quad \quad  + \ B ( 35 \eta^{4} - 30 \eta^{2} + 3)], \label{eq:4.23}  \\
\Phi_{3}(\xi,\eta) &= - \frac{MG}{a} [ \cot^{-1} \xi + \frac{10 A}{6} (3
\eta^{2} - 1)  \nonumber \\
	& \quad \quad + \ \frac{21 B}{11} (35 \eta^{4} - 30 \eta^{2} + 3) 
\nonumber	\\
&  \quad \quad  + \ C (231 \eta^{6} - 315 \eta^{4} + 105 \eta^{2} - 5) ],
\label{eq:4.24}  
\end{align}\end{subequations}
where
\begin{subequations}\begin{align}
A &= \frac{1}{4} [(3\xi^{2} + 1) \cot^{-1} \xi - 3 \xi ],
\\
B &= \frac{3}{448} [ (35 \xi^{4} + 30 \xi^{2} + 3) \cot^{-1} \xi
- 35 \xi^{3} - \frac{55}{3} \xi ], \\
C &= \frac{5}{8448} [ (231 \xi^{6} + 315 \xi^{4} + 105 \xi^{2} +
5) \cot^{-1} \xi  \nonumber \\
& \quad \quad - 231 \xi^{5} - 238 \xi^{3} - \frac{231}{5} \xi ],
\end{align}\end{subequations}
with similar, but more involved, expressions for greater values of $m$.

By taking $\xi = 0$ at the above expressions, we can obtain the value of the
potential at the plane of the disk. Now, is easy to see that, for all  the
members of the family, we will obtain finite values for these quantities. In
particular, for the first member of the family, the potential at the disk is
given by
\begin{equation}
\Phi_1 (R,0) = \frac{3 \pi M G}{8 a^3} R^2
\end{equation}
for $R \leq a$, and this expression is completely equivalent to the
corresponding expression in \cite{KAL}.

In the same way, we obtain for the surface mass densities of the first three
members of the family the expressions
\begin{subequations}\begin{align}
\Sigma_{1} & = \frac{3M}{2\pi a^{2}} (1 - \tilde R^{2})^{1/2},
\label{eq:densi2} \\
\Sigma_{2} & = \frac{5M}{2\pi a^{2}} (1 - \tilde R^{2})^{3/2},
\label{eq:densi3} \\
\Sigma_{3} & = \frac{7M}{2\pi a^{2}} (1 - \tilde R^{2})^{5/2},
\label{eq:densi4}
\end{align}\end{subequations} 
and for the corresponding circular velocities the expressions
\begin{subequations}\begin{align}
V_1^{2} &= \frac{3\pi MG}{4a} \tilde R^{2}, \label{eq:4.43} \\
V_2^{2} &= \frac{15\pi MG}{32a} \tilde R^{2} (4 - 3\tilde R^{2}),
\label{eq:4.44} \\
V_3^{2} &= \frac{105\pi MG}{256a} \tilde R^{2} (5\tilde R^{4} - 12\tilde R^{2}
+ 8), \label{eq:4.45}
\end{align}\end{subequations}
where has been introduced the dimensionless radial variable ${\tilde R} = R/a$.

\begin{figure}
\centering
\epsfig{width=2.95in,file=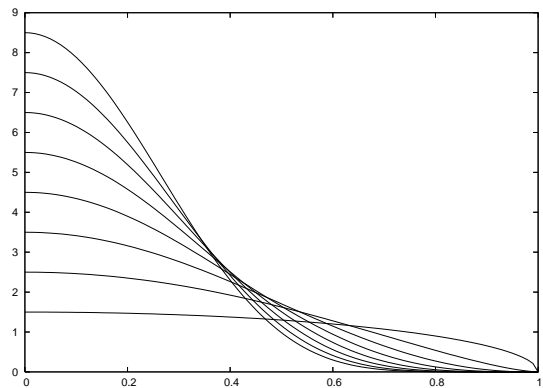}
\caption{Surface density $\tilde\Sigma_m$ as function of $\tilde R$ for
generalized Kalnajs disks models with $m = 1$ (bottom curve) until $m = 8$
(upper curve).}\label{fig:dens}
\end{figure}

In order to graphically illustrate the behavior of the different particular
models, we first introduce the dimensionless surface density of the disks,
defined as
\begin{equation}
\Sigma_{m}(R) = \frac{M}{\pi a^{2}} \tilde \Sigma_{m}(\tilde R),
\end{equation}
for $0 \leq {\tilde R} \leq 1$. So, in Fig. \ref{fig:dens} are depicted the
dimensionless surface mass densities $\tilde \Sigma_{m}$ for the models
corresponding to $m = 1, ... , 8$. As we can see, the disks with higher values
of $m$ present a mass distribution that is more concentrated at the center and
less at the edge. Accordingly, these disks can then be considered as
appropriated models of galaxies with a central bulb.  

\begin{figure}
\centering
\epsfig{width=2.95in,file=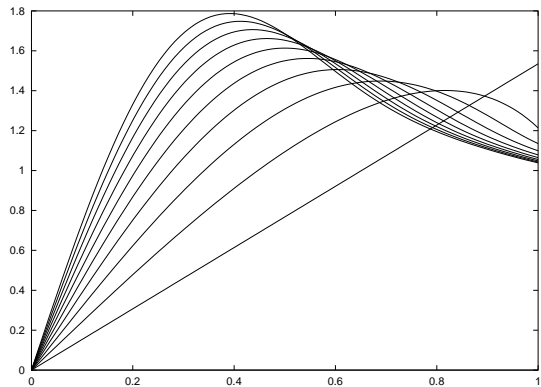}
\caption{Rotation curves $\tilde V_m$ as functions of $\tilde R$ for
generalized Kalnajs disks models with $m=1$ (right line) until  $m = 10$ (upper
curve).}\label{fig:vel1}
\end{figure}

Now, in order to graphically illustrate the behavior of the circular velocities
or rotation curves, we introduce the dimensionless quantity
\begin{equation}
V_m (R) = \sqrt{\frac{MG}{a}} \ \tilde V_m (\tilde R).
\end{equation}
We plot, in Fig. \ref{fig:vel1} the dimensionless rotation curves for the
models corresponding to $m = 1, ... , 10$. The circular velcotiy corresponding
to $m = 1$ is proportional to the radius, representing thus a uniformly
rotating disk. On the other hand, for $m > 1$, the circular velocity increases
from a value of zero at the center of the disks, until it attain a maximum at a
critical radius and then decreases to a finite value at the edge of the disk.
Also we  can see that the value of the critical radius decreases as the value
of $m$  increases.

\section{Concluding Remarks}

We presented an infinite family of axially symmetric thin disks of finite
radius obtained by means of a particularization of the Hunter method. The disk
models so obtained are generalizations of the well known Kalnajs disk, which
corresponds to the first member of the family. The particularization of the
Hunter model was obtained by requiring that the surface density was a
monotonically decreasing function of the radius, with a maximum at the center
of the disk and vanishing at the edge, in such a way that the mass distribution
of the  higher members of the family were more concentrated  at the center. 

We also analyzed the rotation curves of the models and we find for the first
member of the family, the Kalnajs disk, a circular velocity proportional to the
radius, representing thus a uniformly rotating disk, whereas for the other
members of the family the circular velocity increases from a value of zero at
the center of the disks until reach a maximum at a critical radius and then 
decreases to a finite value at  the edge of the disk. Also we find that the
value of the critical radius decreases as the value of $m$ increases.

We believe that the obtained thin disk models have some remarkable properties
and so they can be considered as appropriated realistic flat galaxy models, in
particular if the  superposition of these thin disks with appropriated  halo
distributions (\cite{BT}) is considered. We are now considering some research
in this direction. We are now also working in the non axially symmetric
generalization of the here presented disks models and also in the obtention of
the relativistic generalization of them for the axially symmetric case. 

\section*{Acknowledgments}

The authors want to thank the financial support from COLCIENCIAS, Colombia.

\end{document}